\documentclass[preprint,unsortedaddress,amsmath,amssymb,superscriptaddress]{revtex4-1}

\usepackage{graphicx}
\usepackage{bm}

\usepackage{newfloat}

\usepackage{hyperref}
\hypersetup{
    colorlinks=true, 
    linktoc=all,     
    linkcolor=blue,
    citecolor=blue,
    urlcolor=black,
    breaklinks=true
    }  
\usepackage[all]{hypcap}

\newcommand{\sub}[2]{$#1_\text{#2}$}
\renewcommand{\baselinestretch}{1.3}

\usepackage{lineno}

\begin{document}

\title{Pseudogap and Fermi arc induced by Fermi surface nesting in a centrosymmetric skyrmion magnet}

\author{Yuyang~Dong}
\author{Yuto~Kinoshita}
\affiliation{Institute for Solid State Physics, The University of Tokyo, Kashiwa, Chiba 277-8581, Japan}

\author{Masayuki~Ochi}
\affiliation{Department of Physics, Osaka University, Toyonaka, Osaka 560-0043, Japan}
\affiliation{Forefront Research Center, Osaka University, Toyonaka, Osaka 560-0043, Japan}

\author{Ryu~Nakachi}
\author{Ryuji~Higashinaka}
\affiliation{Department of Physics, Tokyo Metropolitan University, Tokyo 192-0397, Japan}

\author{Satoru~Hayami}
\affiliation{Department of Physics, Hokkaido University, Sapporo 060-0810, Japan}

\author{Yuxuan~Wan}
\author{Yosuke~Arai}
\author{Soonsang~Huh}
\affiliation{Institute for Solid State Physics, The University of Tokyo, Kashiwa, Chiba 277-8581, Japan}

\author{Makoto~Hashimoto}
\author{Donghui~Lu}
\affiliation{Stanford Synchrotron Radiation Lightsource, SLAC National Accelerator Laboratory, Menlo Park, CA 94025, USA}

\author{Masashi~Tokunaga}
\affiliation{Institute for Solid State Physics, The University of Tokyo, Kashiwa, Chiba 277-8581, Japan}
\affiliation{Trans-scale Quantum Science Institute, The University of Tokyo, Tokyo 113-0033, Japan}

\author{Yuji~Aoki}
\author{Tatsuma~D.~Matsuda}
\affiliation{Department of Physics, Tokyo Metropolitan University, Tokyo 192-0397, Japan}

\author{Takeshi~Kondo}
\email{kondo1215@issp.u-tokyo.ac.jp}
\affiliation{Institute for Solid State Physics, The University of Tokyo, Kashiwa, Chiba 277-8581, Japan}
\affiliation{Trans-scale Quantum Science Institute, The University of Tokyo, Tokyo 113-0033, Japan}
\date{\today}

\maketitle

{\bf Skyrmions in noncentrosymmetric materials are believed to occur due to the Dzyaloshinskii-Moriya interaction. By contrast,
the skyrmion formation mechanism in centrosymmetric materials remains elusive. Here, we reveal the intrinsic electronic structure of the centrosymmetric GdRu$_2$Si$_2$ by selectively measuring magnetic domains using angle-resolved photoemission spectroscopy (ARPES). We found robust Fermi surface (FS) nesting, consistent with the magnetic modulation
\emph{\textbf{q}}-vector detected by the previous resonant x-ray scattering measurements. The pseudogap opens at the nested FS portions, which vary for different magnetic domains. The anomalous pseudogap disconnects the FS to generate Fermi
arcs with twofold symmetry. These results indicate that the Ruderman-Kittel-Kasuya-Yosida (RKKY) interaction plays a
decisive role in generating the screw spin modulation responsible for the skyrmion formation in GdRu$_2$Si$_2$.
Furthermore, we demonstrate the flexible nature of magnetism in GdRu$_2$Si$_2$ by manipulating magnetic domains with magnetic field and temperature cyclings, providing potential future applications for data storage and processing devices.}\\

Magnetic skyrmions are particles with spiral spin textures. 
Since their discovery about a decade ago, magnetic skyrmions have attracted enormous research interest because they exhibit exotic phenomena and are promising for spintronic device applications~\cite{neubauer2009a,Yu2010,schulz2012,romming2013,Nagaosa2013}. 
Although most magnetic skyrmion materials are non-centrosymmetric~\cite{muhlbauer2009,seki2012,kezsmarki2015,tokunaga2015},  centrosymmetric systems yielding skyrmion lattice  
have recently been discovered~\cite{Kurumaji2018,Hirschberger2019,Khanh2020,ishiwata2020,gao2020,takagi2022}.
One notable feature of skyrmions in centrosymmetric systems is their tiny size (\textless4 nm), much smaller than that in non-centrosymmetric systems (several tens of nanometers)~\cite{Tokura2021}.
This smaller skyrmion size can lead to more robust electrodynamic responses, which is desirable for high-density data storage and processing devices~\cite{takashima2016,fert2017,psaroudaki2017,yokouchi2020,psaroudaki2021}.
  
GdRu$_2$Si$_2$ is known to form the smallest size ($\sim$1.9 nm) of skyrmions among all materials hosting skyrmions. Whereas the formation mechanism of skyrmions in non-centrosymmetric materials is widely agreed to be the Dzyaloshinskii-Moriya interaction, it remains a challenge to clarify the corresponding mechanism in the centrosymmetric materials~\cite{Kurumaji2018,Hirschberger2019,Khanh2020}. Several theoretical works have proposed geometrical frustration~\cite{okubo2012,leonov2015} or orbital frustration~\cite{Nomoto2020} as key for the skyrmion lattice state; in some others, it is suggested that RKKY interaction~\cite{Inosov2009,wang2020,mitsumoto2021,Bouaziz2022,Dong_PRL} and higher-order spin interactions are required~\cite{ozawa2017,Hayami2017,Hayami2021a}. 
These mechanisms would significantly modify the electronic structure. Quantum oscillation experiments are a good measure to observe the Fermi surface (FS)~\cite{matsuyama2023}. However, they have been inconclusive in distinguishing between those proposals. The investigation by ARPES is also not straightforward 
because structural terminations on the sample surface and magnetic domains can greatly complicate the signals~\cite{Eremeev_ARPES}.

In this study, we reveal the electronic structure of GdRu$_2$Si$_2$ 
using an ARPES technique that solves the difficulties caused by both surface terminations and magnetic domains. 
Magnetic phases of GdRu$_2$Si$_2$ have previously been studied by RXS and scanning tunneling microscope (STM) experiments~\cite{Khanh2020,Yasui2020,Khanh2022}. The emergence of a topologically non-trivial skyrmion lattice state (Phase II) requires both low temperatures (below 20 K) and a magnetic field (within a narrow range above 2 T).  
A magnetic field is not compatible with ARPES measurements, so we investigate the electronic structure only of the ground state at a zero magnetic field (Phase I). Nevertheless, it is crucial for elucidating the skyrmion mechanism because the ground and skyrmion phases (Phase I and Phase ll, respectively) share similar magnetic modulation \emph{\textbf{q}}-vectors:  \sub{\emph{\textbf{q}}}{1} $\sim$ (0.22, 0, 0) and \sub{\emph{\textbf{q}}}{2} $\sim$ (0, 0.22, 0). Here, 0.22 r.l.u. corresponds to about 0.33 \AA$^{-1}$. Note that the skyrmion lattice is the superposition of spiral spin textures represented by multiple \emph{\textbf{q}}-vectors~\cite{Tokura2021,DoubleQ_PRB}. 
The formation mechanism of these \emph{\textbf{q}}-vectors should be reflected in the electronic structure of the ground state, and thus its direct measurements by ARPES are critical.\\

\begin{figure}[!ht]
    \centering
    \includegraphics[scale=1]{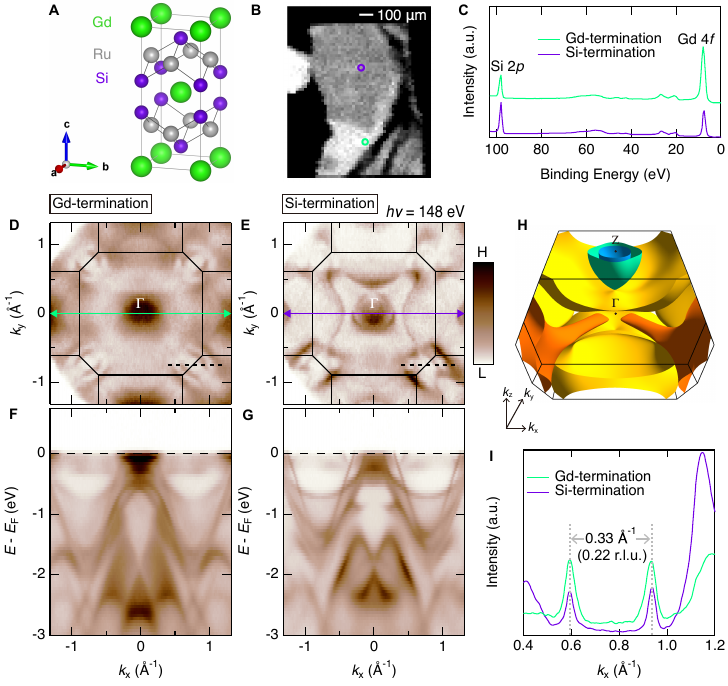}
    \caption{
        \textbf{Surface terminations of GdRu$_2$Si$_2$ and their electronic structures.} 
        (\textbf{A}) Crystal structure of GdRu$_2$Si$_2$. 
         (\textbf{B}) Spatial mapping of photoemission intensity around \sub{E}{F} on the cleaved surface, obtained by scanning the beam spot every 20 $\mu$m. The spot size of the synchrotron light is around 40 $\times$ 10 $\mu$m$^2$. 
 (\textbf{C}) Core level measurements at 2 spots marked by green and purple circles in (B). (\textbf{D},\textbf{E}) Fermi surface mappings of Gd-termination (D) and Si-termination (E) measured at 10 K with 148 eV photons corresponding to \sub{k}{z} = $20 \pi/c$ at the $\Gamma$ point.
For estimating the \sub{k}{z} value, we used the inner potential \sub{V}{0} of 17.9 eV determined from the band dispersion along \sub{k}{z}, which is obtained by changing photon energies in ARPES. The solid black lines denote the bulk Brillouin zone. (\textbf{F},\textbf{G}) Band dispersion maps of Gd-termination and Si-termination, respectively, measured along momentum cuts crossing the $\Gamma$ point (green and purple arrows in (D,E). (\textbf{H}) Calculated FS for the bulk state. (\textbf{I}) Momentum distribution curves (MDCs) along the dashed black lines in (D,E). The dimension arrow represents the peak-to-peak distance, which agrees between the two MDCs.}
    \label{fig:fig1}
\end{figure}

\noindent\textbf{Extracting the common electronic structure of surface terminations}\\
We selectively observe Si-termination and Gd-termination on the cleaved crystal surface [(001) plane] of GdRu$_2$Si$_2$ (Fig.~\ref{fig:fig1}A) by ARPES using synchrotron light with a small beam spot ($\sim$ 40 $\times$ 10 $\mu$m$^2$).
Figure~\ref{fig:fig1}B maps the photoemission intensity near the Fermi energy (\sub{E}{F}). Here, we use the resonant photon energy of 148 eV~\cite{Mishra1998}, which improves contrast in the intensity map. This photon energy also corresponds to the \sub{k}{z} value of $20\pi/c$ at $\Gamma$; here, $c$ is the lattice constant along the  $c$ axis.
A clear contrast of bright and dark distinguishes two areas corresponding either to Gd-termination or Si-termination. The core-level photoemission is performed (Fig.~\ref{fig:fig1}C) for these two areas: the green spot (bright area) and the purple spot (dark area) in Fig.~\ref{fig:fig1}B. The intensity of the Si 2$p$ peak is lower than that of the Gd 4$f$ peak at the green spot. Inversely, the former is higher than the latter at the purple spot. This identifies the green and purple spots as Gd-termination and Si-termination, respectively. We further confirm this by spatially mapping the core level intensities (Fig. S1). 

Next, we conduct ARPES measurements for these two terminations separately.  
The Si-termination (Fig.~\ref{fig:fig1}E) exhibits two circle-shaped FS sheets within a large windmill-shaped FS sheet centered at $\Gamma$. In contrast, these features are not observed for the Gd-termination (Fig.~\ref{fig:fig1}D). Instead, much higher intensities are detected around $\Gamma$. 
Figure~\ref{fig:fig1}F and \ref{fig:fig1}G show the band dispersions of Gd-termination and Si-termination, respectively, passing through $\Gamma$. The overall difference between the two is apparent; for instance, there are some sharp bands near \sub{E}{F} in Si-termination (Fig.~\ref{fig:fig1}G), whereas these bands are missing in Gd-termination (Fig.~\ref{fig:fig1}F). These differences are reproduced by our slab calculations (Fig. S2). 

For a study of the skyrmion mechanism, it is crucial to extract the intrinsic bulk information from these data. 
As a reference, the bulk-state FS calculated with density functional theory (DFT) is displayed in Fig.~\ref{fig:fig1}H.
The FS observed by ARPES is composed of surface and bulk signals. Therefore, the common part of FS observed from Gd-termination and Si-termination should be coming from the intrinsic bulk state. One noticeable common feature is the parallel FS observed at the corner of the Brillouin zone (BZ) in both terminations. Figure~\ref{fig:fig1}I overlays the momentum distribution curves (MDCs) at \sub{E}{F} crossing the BZ corner (dotted lines in Fig.~\ref{fig:fig1}, D and E) for the two terminations. A good agreement is obtained in the peak positions, validating that the parallel FS corresponds to the bulk state. We focus our study on this part of the FS.\\

\begin{figure}[!ht]
    \centering
    \noindent\makebox[\textwidth]{
        \includegraphics[scale=0.9]{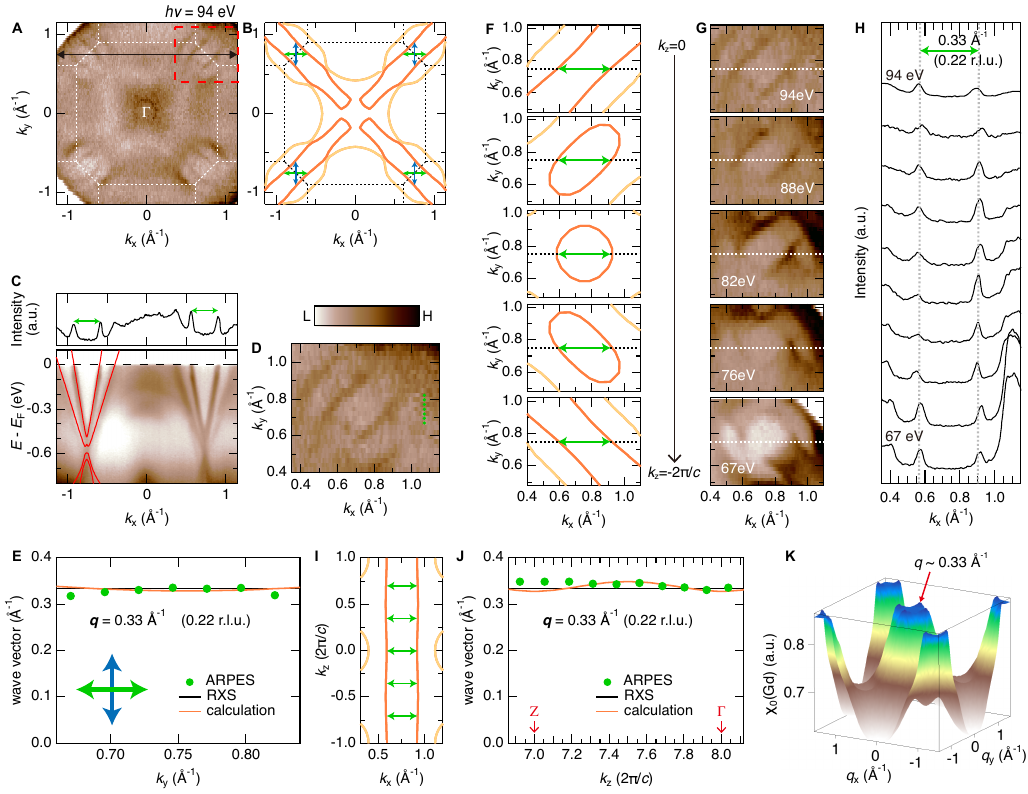}
    }
    \caption{
        \textbf{Fermi surface nesting in the 3D momentum space.} 
        (\textbf{A}) The Fermi surface mapping for Gd-termination measured at 10 K with 94 eV photons corresponding to  \sub{k}{z} = 0. 
        (\textbf{B}) Calculated FS at \sub{k}{z} = 0. The dashed white and black lines in (A) and (B), respectively, denote the Brillouin zone. The green and blue arrows indicate the nesting wave vectors parallel to the \sub{k}{x} and \sub{k}{y} directions, respectively. 
        (\textbf{C}) Band dispersion map connecting the corners of BZ (black arrow in (A)); top: the corresponding momentum distribution curve (MDC) at the Fermi level. The DFT bulk band calculations (red lines) are overlayed on the ARPES dispersion. The length of the nesting wave vector is estimated from the distance between MDC peaks (green arrows). 
        (\textbf{D}) Magnified FS of the red rectangle region in (A). 
        (\textbf{E}) The length of the nesting wave vector along the \sub{k}{x} direction estimated by ARPES (green filled circles) and by calculations (red line), against the \sub{k}{y} positions. The \sub{k}{y} positions for the ARPES results are marked by the green dots in (D). The \emph{\textbf{q}}-vector length of magnetic modulation reported by the RXS study is also overlayed (black line) in (E). 
        (\textbf{F},\textbf{G}) Evolution of the nested in-plane FS with varying \sub{k}{z} (from 0 to $-2\pi/c$), demonstrated by calculation and ARPES, respectively. The values of \sub{k}{z} are changed from $16\pi/c$ (equal to 0) to $14\pi/c$ (equal to $-2\pi/c$) by sweeping photon energy from 94 eV to 67 eV. Note that the period of BZ along the $k_z$ direction is $4\pi/c$ because GdRu$_2$Su$_2$ has a body-centered tetragonal structure.  
        (\textbf{H}) MDCs of the nested FS along the white dotted lines in (G), obtained at different photon energies covering \sub{k}{z}s from 0 to $-2\pi/c$. 
The green arrow connects peak to peak in MDC, which corresponds to 
the length of the nesting wave vector. 
(\textbf{I}) \sub{k}{z} dependence of the nested part of FS along \sub{k}{x} crossing the BZ corner. The green arrows are the same as those in (F). 
(\textbf{J}) As in (E), but for the FS nesting along the \sub{k}{z} direction. 
(\textbf{K}) The Lindhard function estimated for the Gd orbital component $\chi_0$(Gd) from the three-dimensional DFT bulk band structure (Fig. S13 in detail).}
    \label{fig:fig2}
\end{figure}

\noindent\textbf{Fermi surface nesting}\\
We compare side by side the in-plane FS map at \sub{k}{z} = 0 by ARPES with 94 eV photons (Fig.~\ref{fig:fig2}A) and DFT (Fig.~\ref{fig:fig2}B). The agreement between the two is excellent, particularly in the parallel segments around the BZ corner. 
We find robust FS nesting which matches the \emph{\textbf{q}}-vectors of magnetic modulation reported (green and blue arrows in Fig.~\ref{fig:fig2}B). 
In Fig.~\ref{fig:fig2}C, we plot an ARPES band dispersion measured along a momentum cut connecting two BZ corners (black arrow in Fig.~\ref{fig:fig2}A). In the top panel, the MDC at \sub{E}{F} is extracted. 
Sharp spectral peaks pointing to the \sub{k}{F} locations allow us to determine the nesting vector (green arrows in Fig.~\ref{fig:fig2}C). 
Similarly, we observe band dispersions along \sub{k}{x} at different \sub{k}{y} positions marked by green circles in the zoomed FS map (Fig.~\ref{fig:fig2}D) of the red rectangular region in Fig.~\ref{fig:fig2}A.
In Fig.~\ref{fig:fig2}E, the nesting wave vectors determined by ARPES (green circles) are compared with those of our DFT calculations (red line) and the lengths of \emph{\textbf{q}}-vectors reported by RXS measurements~\cite{Khanh2020}.
These three almost perfectly match each other at \emph{\textbf{q}}-vector of $\sim$0.33 \AA$^{-1}$~along \sub{k}{x}. We also note that the ARPES bands almost perfectly match the DFT bulk band (red lines in Fig.~\ref{fig:fig2}C and Fig. S12). 

Next, we examine the FS nesting along the \sub{k}{z} direction. The \sub{k}{z} value of the in-plane FS observed can be controlled by varying photon energies. In Fig.~\ref{fig:fig2}F and \ref{fig:fig2}G, we plot in-plane FS maps around the zone corner 
measured at photon energies from 94 eV to 67 eV covering the entire BZ along \sub{k}{z}. The parallel FS is rotated by 90$^{\circ}$ when the observed \sub{k}{z} is shifted by half the BZ  (Fig.~\ref{fig:fig2}F and  Fig. S5), confirming that the ARPES signals indeed correspond to the bulk state. 
The variation may worsen the overall nesting condition.
However, the FS nesting of the same \emph{\textbf{q}}-vector is valid in the entire BZ along \sub{k}{z}, as long as we focus on the center cut of the nested segments (dashed white lines in Fig.~\ref{fig:fig2}G). 
This is more clearly demonstrated in Fig.~\ref{fig:fig2}H by plotting the MDCs covering the entire BZ along \sub{k}{z}. 
Our DFT calculations reproduce the FS nesting (Fig.~\ref{fig:fig2}I and Fig. S4). These results along \sub{k}{z} are summarized in Fig.~\ref{fig:fig2}J; all three of the nesting wave vectors determined by ARPES and DFT, and \emph{\textbf{q}}-vector by RXS agree with each other.
Our results, therefore, evidence the robust FS nesting in both the in-plane and out-of-plane directions, indicating the presence of RKKY interaction between the 
magnetic moments associated with FS nesting. 
To justify this conclusion further, we calculated the Lindhard function ($\chi_0$) from the three-dimensional DFT bulk band structure (Fig. S13). 
We found that a peak for the nesting vector ($\sim$ 0.33 \AA$^{-1}$) emerges when $\chi_0$(\emph{\textbf{q}}) is calculated for the Gd orbital component (Fig.~\ref{fig:fig2}K and Fig. S13D). It is compatible with the RKKY interaction scenario, which expects the coupling between Gd $5d$ and Gd $4f$ to play the main role in the magnetism of GdRu$_2$Si$_2$.\\

\begin{figure}[!ht]
    \centering
    \noindent\makebox[\textwidth]{
        \includegraphics[scale=1]{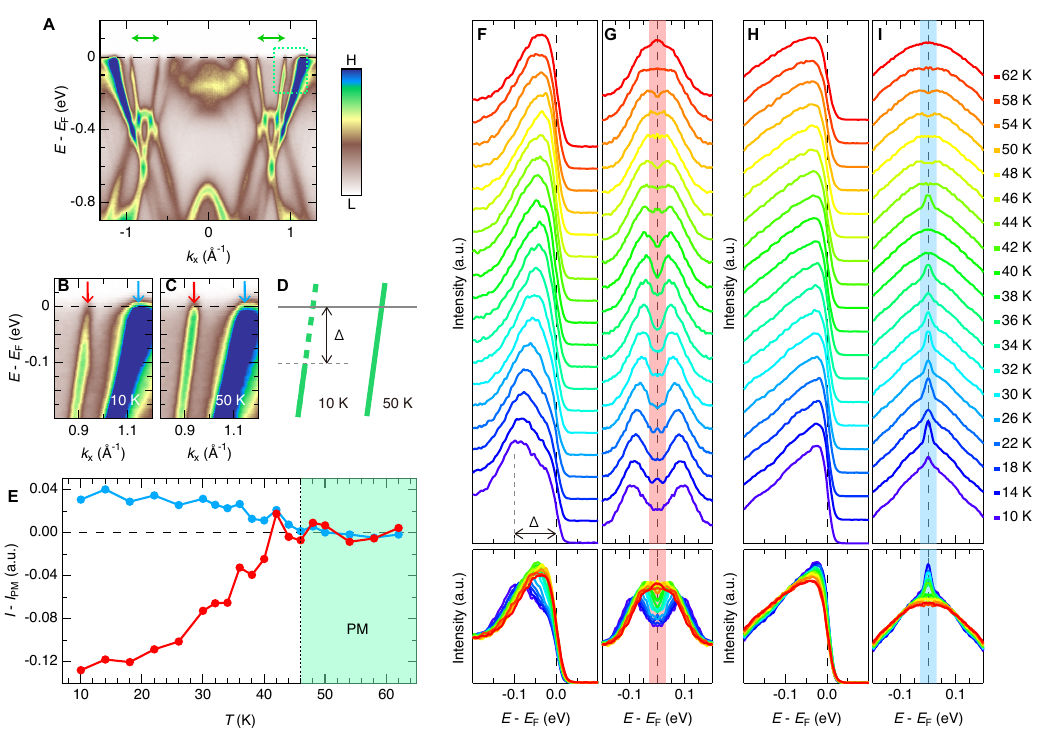}
    }
    \caption{
\textbf{The pseudogap at the nested FS and its temperature evolution.} (\textbf{A}) Band dispersion map of Si-termination along a momentum cut connecting the Brillouin zone corners, which capture the nested parts of FS. The used photon energy is 148 eV, corresponding to \sub{k}{z} = 0. The green arrows indicate the FS nesting wave vector. 
(\textbf{B},\textbf{C}) The magnified image of band dispersions in the green rectangle area of (A) measured below ($T$ = 10 K) and above ($T$ = 50 K) the magnetic transition temperature (\sub{T}{N} = 46 K), respectively. 
The red and blue arrows point to the nesting and non-nesting bands, respectively. (\textbf{D}) Schematic illustration of the nesting band below and above \sub{T}{N} with and without the pseudogap, respectively. 
(\textbf{E}) The temperature evolution of spectral weight near \sub{E}{F} for the nesting (red dots) and non-nesting (blue dots) bands, respectively; each is spectral intensity in (G) and (I) integrated within the red and blue hatched region ($\pm$ 30 meV). 
(\textbf{F},\textbf{H}) The energy distribution curves (EDCs) at \sub{k}{F} (red and blue arrows in (B) and (C)) measured at various temperatures for the nesting and non-nesting band, respectively. In the top and bottom panels, the EDCs are plotted with and without an offset, respectively. 
(\textbf{G},\textbf{I})  EDCs in (F) and (H), respectively, symmetrized about \sub{E}{F}.} 
    \label{fig:fig3}
\end{figure}

\noindent\textbf{Pseudogap associated with FS nesting}\\
Temperature-dependent behavior of the electronic structure is investigated across the N\'eel temperature (\sub{T}{N}) of 46 K. Figure~\ref{fig:fig3}A shows the band dispersion across the nested FS at a low temperature (10 K). Here, we use 148 eV photons, which correspond to \sub{k}{z} at $\Gamma$. The magnified band map in Fig.~\ref{fig:fig3}B (a green-dotted rectangular region in Fig.~\ref{fig:fig3}A) exhibits two band dispersions. The band on the left (red arrow) participates in the FS nesting, whereas the one on the right (blue arrow) does not. Interestingly, we find a loss of spectral weight near \sub{E}{F} only in the nesting band due to the opening of a pseudogap (Fig.~\ref{fig:fig3}D). Notably, the gap is not a full gap; the spectral weight gradually decreases toward \sub{E}{F}, and some amount is left at the \sub{E}{F} even at low temperatures deep below \sub{T}{N}. 
This pseudogap feature is reminiscent of that in high-$T_c$ cuprate superconductors. However, the correlation effects are
less pronounced, and the spectra are sharper in the present case. Consequently, the intensity dispersion in the color-scale map assumes a penpoint-like shape near \sub{E}{F} (Fig.~\ref{fig:fig3}B). The pseudogap is confirmed to close at a high temperature ($T=50$ K) above \sub{T}{N} (Fig.~\ref{fig:fig3}B). In Fig. S8, we demonstrate that the pseudogap is opened along the entire nested portion of FS along \sub{k}{z}, which further evidences the direct relationship between the pseudogap and the FS nesting. In passing, we note that the data do not exhibit band splitting below \sub{T}{N}. If there were band splitting, the nesting wave vector would change between low and high temperatures; this, however, does not occur. 

In Fig.~\ref{fig:fig3}F, we examine the detailed temperature dependence of energy distribution curves (EDCs) at \sub{k}{F} for the nesting band (red arrow in Fig.~\ref{fig:fig3}, B and C). The pseudogap magnitude is estimated to be about 100 meV at 10 K.
With increasing temperature, the pseudogap decreases and eventually closes (Fig.~\ref{fig:fig3}F). 
This gets clearer by symmetrizing EDCs about \sub{E}{F} (Fig.~\ref{fig:fig3}G). Two peaks merge into one peak, closing the gap, at elevated temperatures. In contrast, the spectra at \sub{k}{F} for the non-nesting band (blue arrow in Fig.~\ref{fig:fig3}, B and C) do not open a gap, and they only show the thermal broadening (Fig.~\ref{fig:fig3}, H and I). 
These results are further examined in Fig.~\ref{fig:fig3}E by extracting the spectral weight around \sub{E}{F} (color-hatched area in Fig.~\ref{fig:fig3}, G and I), which is most sensitive to the gap filling.
The spectral weight of the nesting band (red) increases up to \sub{T}{N}, saturating in the paramagnetic (PM) phase. In contrast, the spectral weight for the non-nesting band (blue) monotonically decreases as temperature increases due to the thermal broadening. The pseudogap is, therefore, closely related to the magnetic order. It also explains the upturn at \sub{T}{N} in the resistivity behavior previously reported~\cite{Samanta2008b} and for our samples (Fig. S6).\\

\begin{figure}[!ht]
    \centering
    \noindent\makebox[\textwidth]{
        \includegraphics[scale=0.9]{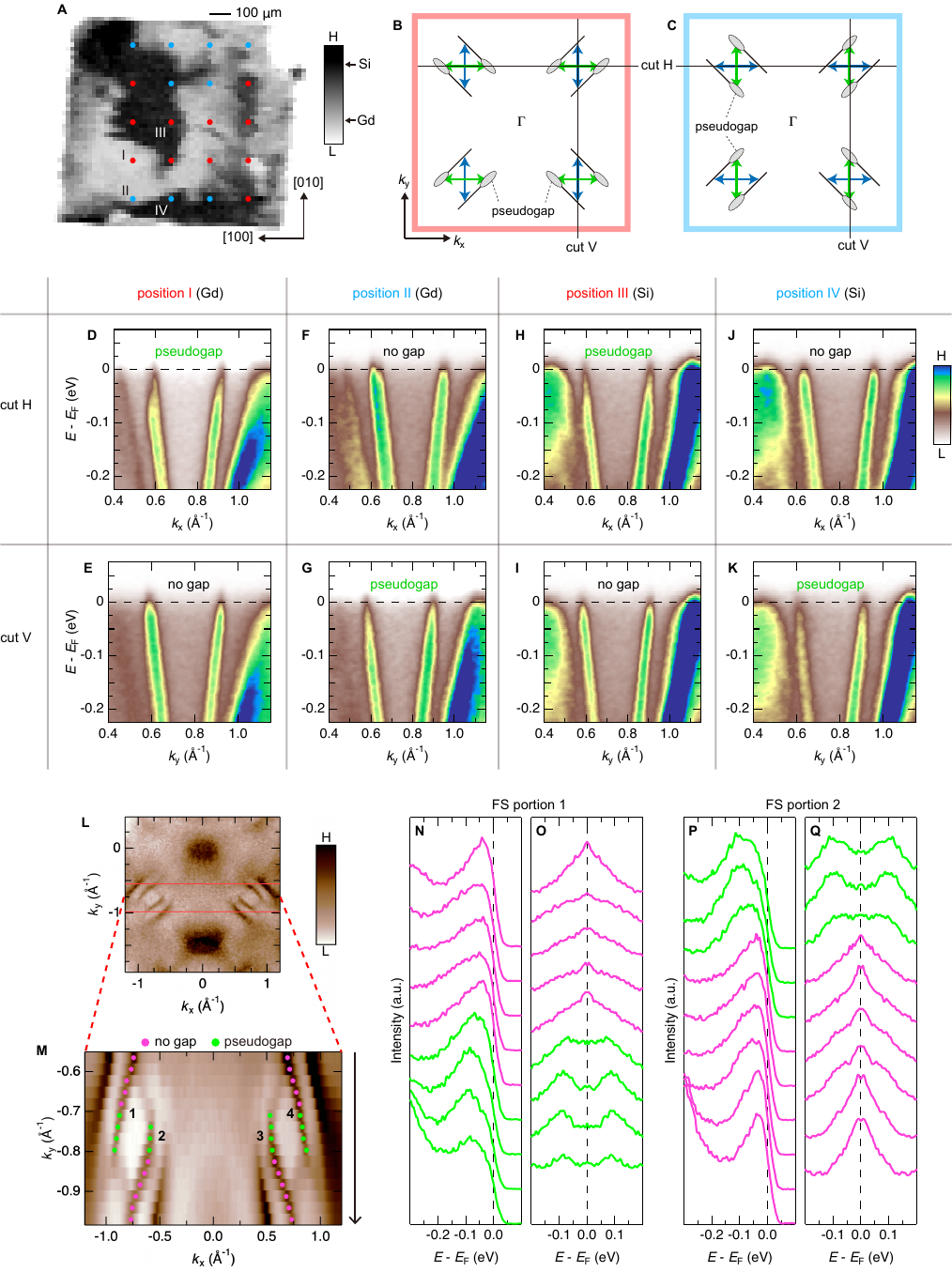}
    }
    \caption{} 
    \label{fig:fig4}
\end{figure}

\noindent\textbf{Magnetic-domain-dependent pseudogap and Fermi arc}\\
We next perform magnetic domain-selective measurements. 
Figure~\ref{fig:fig4}A shows the spatial mapping of photoemission intensity near \sub{E}{F} on the sample surface of GdRu$_2$Si$_2$. One can distinguish Si-termination and Gd-termination from high and low intensities, respectively.
We select 20 spots that are evenly spaced and cover both terminations for the ARPES measurements. 
In Fig.~\ref{fig:fig4}, D-K, we display the nesting bands measured at 10 K along the BZ corners for 4 spots (Position I, II, III, IV, marked in  Fig.~\ref{fig:fig4}A) as examples. 
Here, the left two rows are for Gd-termination, and the right two rows are for Si-termination. For each spot, we measured two different momentum cuts: the horizontal cut along \sub{k}{x} (cut H) and the vertical cut along \sub{k}{y} (cut V), which are separately displayed on the top and low panels. These two cuts were measured at the same experimental geometry for a fair comparison (Section X of~\cite{SM} and Fig. S7). The pseudogap is observed for all the spots. Interestingly, however, we find that only one of the paired two cuts (cut H or cut V) shows the opening of the pseudogap; that is, at 
some spots (Position I, III), the pseudogap opens only on cut H, but at the others (Position II, IV), it opens only on cut V (Fig.~\ref{fig:fig4}, B and C, respectively). 

These data have two implications. One is that the Fermi surface is not closed but arc-like with disconnected portions; this abnormal Fermi arc emerges due to the pseudogap that opens only in partial momentum space.  Second is that the electronic structure breaks the 4-fold symmetry of the crystal structure, becoming a 2-fold symmetry in the ground state (Phase I). This causes two kinds of magnetic domains with the pseudogap opened in different directions. Importantly, both the situation shown in Fig.~\ref{fig:fig4}B (Position I, III) and the one in Fig.~\ref{fig:fig4}C (Position II, IV) can be observed on the same termination (Positions I, II have Gd-termination and Positions III, IV have Si-termination). 
Therefore, the two kinds of magnetic domains are independent of terminations, and they should originate from an intrinsic bulk property. This is more clearly demonstrated in Fig.~\ref{fig:fig4}A by distinguishing all the measured 20 spots with red and blue. The magnetic domains are quite large, and their shapes and locations are independent of the terminations. 

The emergence of the Fermi arc is further validated by investigating the in-plane momentum evolution of the pseudogap. In Fig.~\ref{fig:fig4}M, the red rectangle area of the FS mapping in Fig.~\ref{fig:fig4}L is magnified to focus on the nested FSs.  Here, the Gd-termination is measured, and the magnetic domain observed corresponds to that for Fig.~\ref{fig:fig4}B.
We mark the Fermi momenta \sub{k}{F} with colored dots for the four FS portions 1-4 participating in the FS nesting. 
The EDCs for the FS portions 1 and 2 are extracted in Fig.~\ref{fig:fig4}, N and P, respectively (see Fig. S10 and Fig. S9 for more detail). 
These EDCs are symmetrized in Fig.~\ref{fig:fig4}, O and Q to visualize the gap opening (two peaks structure; green line) or no gap (one peak structure; magenta line). The obtained results are summarized in Fig.~\ref{fig:fig4}M by distinguishing them by the different colors of dots.
We find that the pseudogap opens only at the \sub{k}{F}s (green dots in Fig.~\ref{fig:fig4}M) that can be connected by the horizontal \emph{\textbf{q}}-vector (see Fig.~\ref{fig:fig4}B). The clear crossover from no gap to pseudogap regions along the Fermi surface manifests that the Fermi arc indeed emerges in the magnetic state of GdRu$_2$Si$_2$. Notably, the momentum locations of the Fermi arc are magnetic-domain dependent.
Here, we emphasize that although the relationship between the FS nesting and antiferromagnetism in solids has been studied for so long in condensed matter physics~\cite{oldTheory}, the pseudogap and the associated emergence of the Fermi arc is unexpected, and they are unique for the skyrmion magnet GdRu$_2$Si$_2$.

The two-fold electronic structure revealed by ARPES is compatible with the anisotropic magnetic double-\emph{\textbf{q}} state observed by RXS. The previous RXS measurements uncovered that the magnetic structure of Phase I at a zero magnetic field is formed by the superposition of the screw spin modulation \sub{\emph{\textbf{q}}}{1}= (0.219, 0, 0) [or equivalent \sub{\emph{\textbf{q}}}{1}= (0, 0.219, 0)] and the sinusoidal spin modulation \sub{\emph{\textbf{q}}}{2}= (0, 0.224, 0) [or equivalent \sub{\emph{\textbf{q}}}{2}= (0.224, 0, 0)]~\cite{Khanh2022}. 
The difference in the length of these two perpendicular \emph{\textbf{q}}-vectors  
is so tiny that it is within the measurement error of the nesting wave vector by ARPES.
Each parallel part of FS can, therefore, be connected by both \sub{\emph{\textbf{q}}}{1} and \sub{\emph{\textbf{q}}}{2}.
Our ARPES results that the pseudogap opens along only one of \sub{k}{x} or \sub{k}{y} direction (green arrows in Fig.~\ref{fig:fig4}, B and C) means that only one of the magnetic modulations (either screw-type \sub{\emph{\textbf{q}}}{1} or sinusoidal-type \sub{\emph{\textbf{q}}}{2}) generates the pseudogap. 

We conjecture that the pseudogap is generated by the screw spin modulation, not the sinusoidal spin modulation. 
The screw-type modulation consists of spin moments with equal lengths, contrasting with the sinusoidal-type modulation consisting of shorter spin moments on average (Fig.~\ref{fig:fig5}H). Therefore, the screw spin modulation should be more strongly coupled with the electronic structure than the sinusoidal spin modulation~\cite{ozawa2016}. The skyrmion lattice is formed as the superposition of two perpendicular screw spin modulations. Hence, our results suggest that the pseudogap opens at a large portion of the nested FS simultaneously connected by two screw-type \emph{\textbf{q}}-vectors of each vertical and horizontal direction, energetically stabilizing the skyrmion state under a magnetic field. 


To investigate the properties of the magnetic domains further, we perform polarizing microscopy measurements. 
Figure~\ref{fig:fig5}, A and C show the spatial mapping of photoemission intensity obtained from two sample pieces (sample \#1 and sample \#2). The red and blue spots mark different magnetic domains distinguished by ARPES via observing the pseudogap which opens at different momenta (Fig. S15). 
For the same sample surfaces after the ARPES measurements, we also took the polarizing microscopy images (Fig.~\ref{fig:fig5}, B and D for sample \#1 and sample \#2, respectively). The magnetic domains determined by ARPES match those measured by the polarizing microscopy.\\

\noindent\textbf{Manipulation of the magnetic domains}\\
Finally, we demonstrate manipulations of the magnetic domains in GdRu$_2$Si$_2$ by a magnetic field and temperature cyclings. 
Figure \ref{fig:fig5}E shows a schematic phase diagram with Fig. \ref{fig:fig5}F-I representing magnetic ordering in each phase, adopted from previous RXS research \cite{Khanh2022}.
Both the magnetic field and temperature cyclings excite the system to a state with a 4-fold symmetry and then take it back to a state with a 2-fold symmetry. For the magnetic field cycling, we used a pulsed field with a pulse width of 4.6 ms. The polarizing microscopy images obtained by a series of experiments are shown in Fig.~\ref{fig:fig5}J-\ref{fig:fig5}M. For all the images, 
 As the first procedure, the temperature was decreased from 50 K (PM) to 10 K (Phase I) without a magnetic field (Fig.~\ref{fig:fig5}J: \textcircled{\small{4}}$\rightarrow$\textcircled{\small{1}}). Two domains (blue and white areas) clearly appear at 10 K in Phase I. 
Secondly, the magnetic state was excited from Phase I to Phase II (skyrmion state) by a pulsed magnetic field, which comes back to Phase I in a short time (Fig.~\ref{fig:fig5}K: \textcircled{\small{1}}$\rightarrow$\textcircled{\small{2}}$\rightarrow$\textcircled{\small{1}}). We find that a large number of domain bubbles appear near the original domain boundary, although the main parts of the domains remain almost unchanged. 
Thirdly, the magnetic state was excited from Phase I to Phase III and then taken back to Phase I again by a pulsed magnetic field (Fig.~\ref{fig:fig5}L: \textcircled{\small{1}}$\rightarrow$\textcircled{\small{3}}$\rightarrow$\textcircled{\small{1}}). 
Much smaller bubble-like domains appear and their distribution is completely random, indicating that the memory of the original domains is erased. Lastly, the temperature is increased from 10 K (Phase I) to 50 K (PM) and then decreased to 10 K (Phase I) again without an external magnetic field (Fig.~\ref{fig:fig5}M: \textcircled{\small{1}}$\rightarrow$\textcircled{\small{4}}$\rightarrow$\textcircled{\small{1}}). Interestingly, the profile and distribution of magnetic domains are perfectly restored to the original ones (Fig.~\ref{fig:fig5}J).


Our results demonstrate that the magnetic domain patterns in Phase I can be easily manipulated by applying magnetic fields, while they are robust against temperature cycling and can be perfectly restored. The dual character of magnetism, which can erase and restore the memorized domain patterns by the magnetic field and temperature cyclings, is a unique feature of GdRu$_2$Si$_2$ and should be tightly related to the skyrmion formation mechanism, perhaps owing to the higher-order spin interactions between different magnetic modulations.



\begin{thebibliography}{10}

\bibitem{neubauer2009a}
A.~Neubauer, {\it et~al.\/}, {\it Physical Review Letters\/} {\bf 102}, 186602
  (2009).

\bibitem{Yu2010}
X.~Z. Yu, {\it et~al.\/}, {\it Nature\/} {\bf 465}, 901 (2010).

\bibitem{schulz2012}
T.~Schulz, {\it et~al.\/}, {\it Nature Physics\/} {\bf 8}, 301 (2012).

\bibitem{romming2013}
N.~Romming, {\it et~al.\/}, {\it Science\/} {\bf 341}, 636 (2013).

\bibitem{Nagaosa2013}
N.~Nagaosa, Y.~Tokura, {\it Nature Nanotechnology\/} {\bf 8}, 899 (2013).

\bibitem{muhlbauer2009}
S.~M{\"u}hlbauer, {\it et~al.\/}, {\it Science\/} {\bf 323}, 915 (2009).

\bibitem{seki2012}
S.~Seki, S.~Ishiwata, Y.~Tokura, {\it Physical Review B\/} {\bf 86}, 060403
  (2012).

\bibitem{kezsmarki2015}
I.~K{\'e}zsm{\'a}rki, {\it et~al.\/}, {\it Nature Materials\/} {\bf 14}, 1116
  (2015).

\bibitem{tokunaga2015}
Y.~Tokunaga, {\it et~al.\/}, {\it Nature Communications\/} {\bf 6}, 7638
  (2015).

\bibitem{Kurumaji2018}
T.~Kurumaji, {\it et~al.\/}, {\it Science\/} {\bf 365}, 914 (2019).

\bibitem{Hirschberger2019}
M.~Hirschberger, {\it et~al.\/}, {\it Nature Communications\/} {\bf 10}, 5831
  (2019).

\bibitem{Khanh2020}
N.~D. Khanh, {\it et~al.\/}, {\it Nature Nanotechnology\/} {\bf 15}, 444
  (2020).

\bibitem{ishiwata2020}
S.~Ishiwata, {\it et~al.\/}, {\it Physical Review B\/} {\bf 101}, 134406
  (2020).

\bibitem{gao2020}
S.~Gao, {\it et~al.\/}, {\it Nature\/} {\bf 586}, 37 (2020).

\bibitem{takagi2022}
R.~Takagi, {\it et~al.\/}, {\it Nature Communications\/} {\bf 13}, 1472 (2022).

\bibitem{Tokura2021}
Y.~Tokura, N.~Kanazawa, {\it Chemical Reviews\/} {\bf 121}, 2857 (2021).

\bibitem{takashima2016}
R.~Takashima, H.~Ishizuka, L.~Balents, {\it Physical Review B\/} {\bf 94},
  134415 (2016).

\bibitem{fert2017}
A.~Fert, N.~Reyren, V.~Cros, {\it Nature Reviews Materials\/} {\bf 2}, 17031
  (2017).

\bibitem{psaroudaki2017}
C.~Psaroudaki, S.~Hoffman, J.~Klinovaja, D.~Loss, {\it Physical Review X\/}
  {\bf 7}, 041045 (2017).

\bibitem{yokouchi2020}
T.~Yokouchi, {\it et~al.\/}, {\it Nature\/} {\bf 586}, 232 (2020).

\bibitem{psaroudaki2021}
C.~Psaroudaki, C.~Panagopoulos, {\it Physical Review Letters\/} {\bf 127},
  067201 (2021).

\bibitem{okubo2012}
T.~Okubo, S.~Chung, H.~Kawamura, {\it Physical Review Letters\/} {\bf 108},
  017206 (2012).

\bibitem{leonov2015}
A.~O. Leonov, M.~Mostovoy, {\it Nature Communications\/} {\bf 6}, 8275 (2015).

\bibitem{Nomoto2020}
T.~Nomoto, T.~Koretsune, R.~Arita, {\it Physical Review Letters\/} {\bf 125},
  117204 (2020).

\bibitem{Inosov2009}
D.~S. Inosov, {\it et~al.\/}, {\it Physical Review Letters\/} {\bf 102}, 046401
  (2009).

\bibitem{wang2020}
Z.~Wang, Y.~Su, S.-Z. Lin, C.~D. Batista, {\it Physical Review Letters\/} {\bf
  124}, 207201 (2020).

\bibitem{mitsumoto2021}
K.~Mitsumoto, H.~Kawamura, {\it Physical Review B\/} {\bf 104}, 184432 (2021).

\bibitem{Bouaziz2022}
J.~Bouaziz, E.~{Mendive-Tapia}, S.~Bl{\"u}gel, J.~B. Staunton, {\it Physical
  Review Letters\/} {\bf 128}, 157206 (2022).

\bibitem{Dong_PRL}
Y.~Dong, {\it et~al.\/}, {\it Phys. Rev. Lett.\/} {\bf 133}, 016401 (2024).

\bibitem{ozawa2017}
R.~Ozawa, S.~Hayami, Y.~Motome, {\it Physical Review Letters\/} {\bf 118},
  147205 (2017).

\bibitem{Hayami2017}
S.~Hayami, R.~Ozawa, Y.~Motome, {\it Physical Review B\/} {\bf 95}, 224424
  (2017).

\bibitem{Hayami2021a}
S.~Hayami, T.~Okubo, Y.~Motome, {\it Nature Communications\/} {\bf 12}, 6927
  (2021).

\bibitem{matsuyama2023}
N.~Matsuyama, {\it et~al.\/}, {\it Physical Review B\/} {\bf 107}, 104421
  (2023).

\bibitem{Eremeev_ARPES}
S.~V. Eremeev, {\it et~al.\/}, {\it Nanoscale Adv.\/} {\bf 5}, 6678 (2023).

\bibitem{Yasui2020}
Y.~Yasui, {\it et~al.\/}, {\it Nature Communications\/} {\bf 11}, 5925 (2020).

\bibitem{Khanh2022}
N.~D. Khanh, {\it et~al.\/}, {\it Advanced Science\/} {\bf 9}, 2105452 (2022).

\bibitem{DoubleQ_PRB}
G.~D.~A. Wood, {\it et~al.\/}, {\it Physical Review B\/} {\bf 107}, L180402
  (2023).

\bibitem{Mishra1998}
S.~R. Mishra, {\it et~al.\/}, {\it Physical Review Letters\/} {\bf 81}, 1306
  (1998).

\bibitem{Samanta2008b}
T.~Samanta, I.~Das, S.~Banerjee, {\it Journal of Applied Physics\/} {\bf 104},
  123901 (2008).

\bibitem{SM}
See Supplementary Materials.

\bibitem{oldTheory}
O.~K. Andersen, T.~L. Loucks, {\it Physical Review\/} {\bf 167}, 551 (1968).

\bibitem{ozawa2016}
R.~Ozawa, {\it et~al.\/}, {\it Journal of the Physical Society of Japan\/} {\bf
  85}, 103703 (2016).

\bibitem{Data_availability}
Y. Dong $et~al$., The dataset for ``Pseudogap and Fermi arc induced by Fermi surface nesting in a centrosymmetric skyrmion magnet". figshare. Dataset. (2025); https://doi.org/10.6084/m9.figshare.28430210.v1

\bibitem{perdew1996}
J.~P. Perdew, K.~Burke, M.~Ernzerhof, {\it Physical Review Letters\/} {\bf 77},
  3865 (1996).

\bibitem{kresse1999}
G.~Kresse, D.~Joubert, {\it Physical Review B\/} {\bf 59}, 1758 (1999).

\bibitem{kresse1993}
G.~Kresse, J.~Hafner, {\it Physical Review B\/} {\bf 47}, 558 (1993).

\bibitem{kresse1994}
G.~Kresse, J.~Hafner, {\it Physical Review B\/} {\bf 49}, 14251 (1994).

\bibitem{kresse1996}
G.~Kresse, J.~Furthm{\"u}ller, {\it Computational Materials Science\/} {\bf 6},
  15 (1996).

\bibitem{kresse1996a}
G.~Kresse, J.~Furthm{\"u}ller, {\it Physical Review B\/} {\bf 54}, 11169
  (1996).

\bibitem{hiebl1983}
K.~Hiebl, C.~Horvath, P.~Rogl, M.~J. Sienko, {\it Journal of Magnetism and
  Magnetic Materials\/} {\bf 37}, 287 (1983).

\bibitem{marzari1997}
N.~Marzari, D.~Vanderbilt, {\it Physical Review B\/} {\bf 56}, 12847 (1997).

\bibitem{souza2001}
I.~Souza, N.~Marzari, D.~Vanderbilt, {\it Physical Review B\/} {\bf 65}, 035109
  (2001).

\bibitem{pizzi2020}
G.~Pizzi, {\it et~al.\/}, {\it Journal of Physics: Condensed Matter\/} {\bf
  32}, 165902 (2020).

\bibitem{kawamura2019}
M.~Kawamura, {\it Computer Physics Communications\/} {\bf 239}, 197 (2019).

\bibitem{katakura2010}
I.~Katakura, {\it et~al.\/}, {\it Review of Scientific Instruments\/} {\bf 81},
  043701 (2010).

\bibitem{TokuraPRB}
T.~Ishikawa, K.~Ookura, Y.~Tokura, {\it Physical Review B\/} {\bf 59}, 8367
  (1999).

\bibitem{kinoshita2022}
Y.~Kinoshita, T.~Miyakawa, X.~Xu, M.~Tokunaga, {\it Review of Scientific
  Instruments\/} {\bf 93}, 073702 (2022).

\bibitem{MommaVESTA}
K.~Momma, F.~Izumi, {\it Journal of Applied Crystallography\/} {\bf 44}, 1272
  (2011).
  
\end{thebibliography}

\bigskip
\textbf{Acknowledgements}\\

We thank K. Kuroda, T. Taniuchi, and  K. Yamagami for fruitful discussion on the results, and also thank T. Yajima for XRD measurements. \textbf{Funding:} Use of the Synchrotron Radiation Lightsource, SLAC National Accelerator Laboratory, is supported by the U.S. Department of Energy, Office of Science, Office of Basic Energy Sciences under Contract No. DE-AC02-76SF00515.
This work was supported by the JSPS KAKENHI (Grants Numbers. JP21H04439, JP19H00651, JP22K03517, JP21H01037, and JP23H04869), by the Asahi Glass Foundation, by MEXT Q-LEAP (Grant No. JPMXS0118068681), by JST PRESTO (Grant No. JPMJPR20L8), by The Murata Science Foundation, by The Mitsubishi Foundation, and by Tokyo Metropolitan Government Advanced Research (Grant Number. H31-1). \textbf{Author contributions:}
T.K. initiated the idea and designed the studies.
T.K. and T.D.M. planned the experiment project. 
Y.D. conducted ARPES experiments and analyzed the data.
Y.D., Y.K., and M.T. took polarizing microscopy images.
Y.W., Y.Arai, S.Huh, M.H., D.L., and T.K. supported the ARPES experiment.
M.O. calculated the theoretical band structure.
R.N., R.H., Y.Aoki, and T.D.M. prepared the single crystals.
M.O. and S.Hayami provided theoretical insights.
Y.D. and T.K. wrote the paper.
All authors discussed the results and commented on the manuscript.
\textbf{Competing interests:} The authors declare no
competing interests.
\textbf{Data and materials availability:} All data are available in the main text, in the supplementary materials, and
on Zenodo.~\cite{Data_availability}\\ 

\textbf{Supplementary Materials}
Materials and Methods
Supplementary Text
Figs. S1 to S17
References (44-60)

\clearpage

\renewcommand{\baselinestretch}{1.05}

\clearpage

\clearpage

\clearpage

\clearpage
\addtocounter{figure}{-1}
\begin{figure} [t!]
  \caption{\textbf{Magnetic-domain-dependent pseudogap.} (\textbf{A}) Spatial mapping of photoemission intensity on the GdRu$_2$Si$_2$ surface. The black and grey areas (strong and weak intensities) indicate Si-termination and Gd-termination, respectively; note that the correspondence relationship between the color contrast (black and grey) and terminations (Si-termination and Gd-termination) is reversed from that of Fig. 1A because the selected energy-momentum region for photoemission intensity mapping is different (see Section III of~\cite{SM} and Fig. S3). Colored dots mark twenty different spots separately measured by ARPES. 
Red and blue represent two kinds of magnetic domains distinguished by the observation of pseudogap. 
(\textbf{B},\textbf{C}) Schematic FS of the two different domains each corresponding to red and blue dot in (A). In the domain of red spots, the pseudogap is observed only along the horizontal cut (cut H). In the domain of blue spots, the pseudogap is observed only along the vertical cut (cut V).
(\textbf{D} to \textbf{K}) Band dispersion maps of 4 different spots marked as I, II, III, and IV in (A). Position I and II are from Gd-termination, whereas Position III and IV are from Si-termination.
The upper panels (D to J) are obtained along cut H, whereas the lower panels (E to K) are along cut V. Whether or not the pseudogap is observed is noted in each panel.
(\textbf{L}) The FS mapping for Gd-termination and the magnetic domain corresponding to (B), measured with 94 eV photons at 10 K.
(\textbf{M}) Magnified FS of the red rectangular area in (L). The momentum evolution of the pseudogap is traced for four FS portions 1-4. 
The \sub{k}{F} points with the pseudogap and no gap are marked by the green and magenta dots, respectively.
(\textbf{N},\textbf{P}) EDCs at \sub{k}{F}s for FS portion 1 and 2, respectively. 
From top to bottom, EDCs are plotted in the order from a large to a small \sub{k}{y} value, as represented by the black arrow in (M).
(\textbf{O},\textbf{Q}) EDCs in (N) and (P) symmetrized about \sub{E}{F}, respectively.
A similar set of data for FS portions 3 and 4 are presented in Fig. S10. 
}
\end{figure}

\begin{figure}[!ht]
    \centering
    \noindent\makebox[\textwidth]{
        \includegraphics[scale=1]{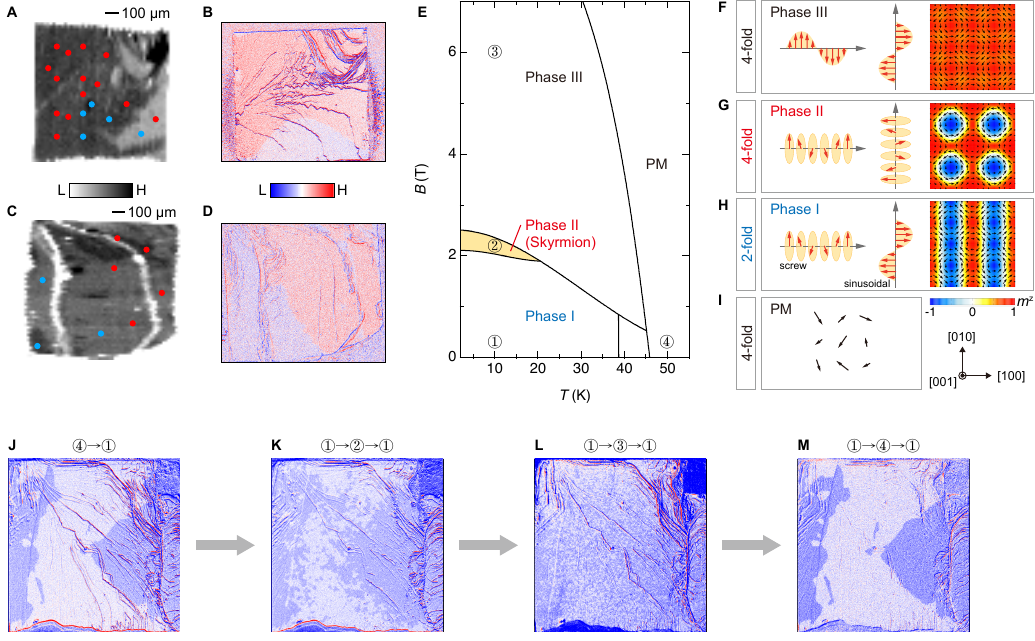}
    }
    \caption{
        \textbf{Manipulation of magnetic domains by field and temperature cyclings.} (\textbf{A},\textbf{C}) Magnetic domains determined by ARPES for sample \#1 and sample \#2, respectively. Spots separately measured by ARPES are marked on each spatial mapping of photoemission intensity around \sub{E}{F}; note that the selected momentum region for this intensity mapping differs from that in Fig. 1A. Red and blue colors indicate different magnetic domains. 
(\textbf{B},\textbf{D}) Magnetic domains visualized by polarizing microscopy. The image plots the intensity difference below and above the N\'eel temperature (10 K and 50 K), displaying 
two kinds of magnetic domains (red and blue areas). 
(\textbf{E}) Magnetic phase diagram of GdRu$_2$Si$_2$~\cite{Khanh2020}. (\textbf{F} to \textbf{I}) Schematics of spin texture for Phase III, Phase II, Phase I, and paramagnetic (PM) phase, respectively~\cite{Khanh2022}. Only Phase I has a 2-fold magnetic structure. (\textbf{J} to \textbf{M}), Magnetic domains observed by polarizing microscopy after different steps in temperature and field cyclings: Cooling down from 50 K to 10 K (J). Excited to Phase II and returning to Phase I (K). Excited to Phase III and returning to Phase I (L). Temperature cycling from 10 K to 50 K and back to 10 K (M). 
For the field cycling, a pulsed magnetic field is used.
All images are measured at 10 K without field; the data in the PM phase at 50 K is subtracted as background.} 
    \label{fig:fig5}
\end{figure}

\clearpage

\end{document}